\documentclass[a4paper,11pt]{article}
\pdfoutput=1 

\usepackage{jheppub} 

\usepackage[T1]{fontenc} 

\newcommand{\Neqfour}{\mathcal{N}=4}
\newcommand{\RthreeSone}{\mathbb{R}^{3}\times\mathrm{S}^{1}}
\newcommand{\Qcdthree}{\mathrm{QCD_{3}^{W}}}

\newcommand{\Ofsquare}{\widehat{\mathcal{O}}_{F^{2}}}
\newcommand{\xbar}{\bar{x}}

\makeatletter 
\def\@fpheader{\relax} %
\title{\boldmath Flux-tubes in confining gauge theories with gravitational dual}

\author{Vikram Vyas}
\affiliation{Ajit Foundation Science Centre\\Bikaner, India}
\emailAdd{vikram@physicsinfield.org}

\abstract{The emergence of flux-tubes as the distance is increased between a
quark and an antiquark is explored in a three-dimensional confining
gauge theory using the gauge/gravity duality. We delineate the intrinsic shape
of the flux-tube corresponding to the classical open string in the
bulk and further explore the fluctuation in its thickness induced
by the fluctuation of the corresponding open string along the radial direction in
the bulk. The relationship between the intrinsic shape of a flux-tube
and its effect on the heavy quark potential is also discussed. 
  }

\begin{document} 
\maketitle
\flushbottom
\section{Introduction}

\label{sec:intro} Various investigations of confining theories like
QCD suggest that a fluctuating flux-tube is formed between a quark
and an anti-quark \cite{Singh_1993,Bali:1994de,Bissey:2009aa}. We
do not completely understand the physics behind the formation of these
flux-tubes, but we expect on the basis of models like the dual-superconductor
model of confinement \cite{PhysRevD.10.4262,tHooft:1979rtg,Mandelstam1980109}
and on fairly general grounds \cite{Wilson:1977nj,FEYNMAN1981479},
that these flux-tubes are not one-dimensional objects like strings
but have a finite intrinsic thickness. Therefore it would be nice
to delineate these flux-tubes and measure their intrinsic thickness
devoid of the quantum broadening, and hopefully obtain a better understanding
of the phenomenon of confinement. One might think that this is the
kind of calculation that can be easily done using the techniques of
lattice gauge theories, unfortunately that is not the case. For the
expectation value of any gauge invariant operator, like the action
density, which we could use to delineate the shape of the flux-tube
would not distinguish between the contribution due to quantum fluctuations
of the flux-tube and the contribution due to the flux-tube itself\footnote{I would like to thank Gunnar Bali for pointing this out to me.},
and it requires a considerable effort to disentangle the two effects
\cite{Cea:2012qw,Cardoso:2013aa,Caselle:2016aa,Cea:2017ocq,Baker:2018hyy}. We will
refer to the profile of the flux distribution delineated by a gauge
invariant operator as the flux-profile.

It is here that the gauge/gravity duality \cite{Maldacena:1997re,Gubser:1998bc,Witten:1998qj}
can provide us with some additional guidance and intuition. For there
are quantum field theories that exhibit confinement and also have
a dual classical gravitational description \cite{Witten:1998qj,Witten:1998zw}
(see \cite{Maldacena:fj} for a brief review of the gauge/gravity
duality that is relevant for the present work.) In such theories the
gravitational description of a pair of quark 
and an anti-quark is an open string living in a higher dimensional
curved space time which starts from the quark and ends at the anti-quark
\cite{Polyakov:1998aa,Maldacena:1998im,Rey:1998ik}, where by quark
we simply mean a source in the fundamental representation of the gauge
group. The open string is in a quantum state which is a superposition
of different configurations, all terminating at the quark and the
antiquark. 
This string is also a source of dilaton field which, through the dictionary
established in \cite{Gubser:1998bc,Witten:1998qj}, induces in the
boundary theory a flux-tube connecting the quark and the anti-quark
\cite{Balasubramanian:aa,Balasubramanian:1998de,Danielsson:1998wt,Callan:2000sf}.
As a result for every configuration of the fluctuating open sting
in the bulk there is a holographic projection in the form of a flux-tube
of varying thickness, the thickness of the flux-tube depending on
the radial position of the corresponding open string in the bulk \cite{Polchinski:2001ju}.
The flux-profile produced in the confining theory by a quark and an
anti-quark can be thought of as superposition of these flux-tubes.

The main aim of the present work is to substantiate these remarks
by explicitly calculating the shape of the flux-tube induced by some
physically motivated configurations of the open string. We will do
this for a confining theory in three dimensions which has a dual classical
gravitational description \cite{Witten:1998qj}. The reason for restricting
to three dimensions is simply that in three dimensions the required
numerical calculations can be done with reasonable precision using
a modest personal computer. All the techniques that we develop can
be easily extended to the the four dimensional case, but would correspondingly
make a larger demand on computational resources. An alternate approach for exploring the flux-profile in a confining gauge theory using gauge/gravity duality which is closer in spirit to the lattice gauge theory calculations is described in \cite{Giataganas:2015yaa}.

The outline of the paper is as follow, in section (\ref{sec:From-dilaton-field})
we review the relationship between the dilaton field sourced by an
open string and the resulting flux-tube in the boundary theory. 
In section (\ref{sec:Framework-for-numerical}) we develop the framework
for numerically calculating the shape of a flux-tube using gauge/gravity
duality. For the model confining theory that we are studying, namely
the Witten's model, the classical open string configuration is responsible
for the linear potential \cite{Kinar:1998vq}, with this as a motivation,
in section (\ref{sec:Shape-of-the-1}) we study the flux-tubes which
are holographic projections of classical open strings. One of the
salient features of the gravitational description of a flux-tube is
that the intrinsic thickness of the flux-tube fluctuates due to the
fluctuation of the open string in the radial direction, we explore
this in section (\ref{subsec:Fluctuation-intrinsicTHickness}). In
studying confining theories using gauge/gravity duality it is natural
to ask if there are any insights that can be carried over to QCD,
since for QCD there is no known dual gravitational description. We
discuss this question in the final section (\ref{sec:Discussion})
of the paper in the context of effective string descriptions of flux-tubes
and the heavy quark potential.

\section{From dilaton fields to flux-profile\label{sec:From-dilaton-field}}

\subsection{The confining theory}

The model confining theory that we will study is $\Neqfour$ super
Yang-Mills theory on $\RthreeSone$. This is an interacting theory
of gluons, four Weyl fermions and six scalar fields. 
All these fields are charged, they belong to the adjoint representation
of $SU(N)$, and thus they interact with each other and themselves.
The gravitational dual of this theory is given by Witten's model \cite{Witten:1998qj}.
This confining three dimensional theory is not the large $N$ limit
of QCD in three dimensions, for example the glue-balls in our confining
theory are not made purely of gluons but will also get contributions
from the fermions and the scalars. This is because the confinement
scale is comparable to the radius $r_{y}$ of $\text{S}^{1}$ (see
for e.g. \cite{Maldacena:fj}). For convenience we will refer to our
confining theory as $\Qcdthree$. 

To study the formation of flux-tubes we introduce a pair of quark
and an anti-quark into the ground state of this theory. For the purpose
of probing the ground state of the theory we can restrict to massive
quarks, 
\begin{equation}
m_{q}>>r_{y}^{-1},\label{eq:heavyQuark}
\end{equation}
and ignore the translational and spin degrees of freedom. The fluctuating
color degrees of freedom of the original ground state, $|\Omega>$,
interact with the fluctuating color degrees of freedom of the test
quarks, and the ground-state get modified \cite{polyakovBook}. We
will refer to the new ground state as 
\begin{equation}
|\Omega>_{q\bar{q}}.\label{eq:VacuumWithQQbar}
\end{equation}
One possible way of delineating the flux profile is to plot the expectation
value of $\mathrm{Tr}F^{2}$, where $F$ is the Yang-Mills field strength
tensor, as a function of the two spatial coordinates, after averaging
over the compact $y$ direction, 
\begin{equation}
\bar{x}=\left(x_{1},x_{2}\right).\label{eq:defXbar}
\end{equation}
In the context of gauge/gravity duality a more convenient operator
is the generalization of $\text{Tr}F^{2}$ to the supersymmetric case,
which is the Lagrangian of the $\Neqfour$ super Yang-Mills theory
on $\RthreeSone$. This operator can be written schematically as (see
for e.g. \cite{ammon2015gauge} for the details) 
\begin{equation}
\mathcal{O}_{F^{2}}\left(x\right)=\frac{1}{4g_{YM}^{2}}\mathrm{Tr}\left\{ F^{2}+\text{ scalar + fermion}\right\} ,\label{eq:LofN=00003D00003D00003D4SYM}
\end{equation}
where $\text{scalar}$ and $\text{femion}$ represent the kinetic
energy term for the scalar and the fermionic degrees of freedom. With
this in mind we define the profile of the flux-tube as 
\begin{equation}
P\left(\xbar\right)=<\Omega|\Ofsquare\left(\xbar\right)|\Omega>_{q\bar{q}},\label{eq:expectationValueOfFsquare}
\end{equation}
where we have disregarded the dependance of this expectation value
on the $y$ coordinate of the compact direction, or more accurately
we will effectively average over the compact $y$ direction.

\subsection{Flux profile using the gauge/gravity duality}

We would like to calculate the flux-profile as defined by (\ref{eq:expectationValueOfFsquare})
using the gauge/gravity duality. In the classical gravitational description,
the boundary state $|\Omega>_{q\bar{q}}$ is described by a quantum
open string living in a specific curved space-time. 
A given configuration of an open string is also a source for dilaton field
which induces a non-zero expectation value for the operator $\Ofsquare$
in the boundary theory. Since the linear confining potential arises
from the classical string configuration, we will focus on the dilaton
field generated by the classical open string and will refer to the
flux-tube induced by it as the classical flux-tube. 


To calculate the dilaton field produced by an open string we will follow \cite{Callan:2000sf}  except that we will be considering open strings in a confining background. We start with the Nambu-Goto action for
an open string, 
\begin{equation}
S_{NG}=\frac{1}{2\pi\alpha'}\int d^{2}\sigma\sqrt{g},\label{eq:NGinStringFrame}
\end{equation}
where $g$ is the determinant of the two-dimensional metric induced
by the five-dimensional bulk metric $G_{mn}$. The metric $G_{mn}$
experienced by the string is often referred to as the metric in the
string frame while the bulk metric 
that provides the gravitational description of $\Qcdthree$ is the
metric $G_{mn}^{E}$ in the Einstein frame\footnote{For a pedagogical discussion of the relationship between the string
frame and the Einstein frame see \cite{Tong:2009aa}.}, 
\begin{equation}
ds^{2}=G_{mn}^{E}dx^{m}dx^{n}=\frac{R^{2}}{z^{2}}\left(dt_{E}^{2}+d\bar{x}^{2}+(1-\frac{z^{4}}{z_{0}^{4}})dy^{2}+(1-\frac{z^{4}}{z_{0}^{4}})^{-1}dz^{2}\right),\label{eq:confingMetricQCD3}
\end{equation}
and the two are related by 
\begin{equation}
G_{mn}^{E}=e^{-\frac{\Phi}{2}}G_{mn},\label{eq:stringFrameToEinsteinFrame}
\end{equation}
where $\Phi\left(z,t_{E},\bar{x},y\right)$ is the dilaton field and
$z$ is the coordinate along the fifth dimension, which we will often
refer to as the radial direction.

The Nambu-Goto action when written in the Einstein frame becomes 
\begin{equation}
S_{NG}=\frac{1}{2\pi\alpha'}\int d^{2}\sigma e^{\frac{\phi}{2}}\sqrt{g_{E}},\label{eq:NGinEframe}
\end{equation}
where $g_{E}$ is the determinant of the world-sheet metric induced
by $G_{MN}^{E}.$ The complete dynamics of the dilaton field in the
presence of an open string is given by the action 
\begin{equation}
S[\Phi]=S_{dilaton}+S_{NG}\label{eq:bulkAction}
\end{equation}
with $S_{dilaton}$ being the action of a free massless scalar field
in the bulk. This action, after Kaluza-Klein reduction over $S_{5}$,
takes the following form 
\begin{equation}
S_{dilaton}=\frac{R^{5}\Omega_{5}}{4\kappa^{2}}\int d^{5}x\sqrt{G_{E}}G_{E}^{mn}\partial_{m}\Phi\partial_{n}\Phi,\label{eq:Sdilation}
\end{equation}
where $R$ is the radius of $\text{S}_{5}$ and $\Omega_{5}$ is the
volume of a unit sphere in five dimensions. In the above equation
we have defined 
\begin{equation}
\kappa^{2}=4\pi G_{10}.\label{eq:defKappa}
\end{equation}
The complete linearized action for the dilaton field in the presence
of an open string is then given by 
\begin{align}
S_{B}[\Phi] & =\frac{1}{4\kappa_{5}^{2}}\int d^{5}x\sqrt{G_{E}}G_{E}^{mn}\partial_{m}\Phi\partial_{n}\Phi+\frac{1}{2\pi\alpha'}\int d^{2}\sigma\sqrt{g_{E}\left(X\right)}\label{eq:linearizedBulkPhiAction}\\
 & +\frac{1}{4\pi\alpha'}\int d^{2}\sigma\Phi\left[X\right]\sqrt{g_{E}\left(X\right)},
\end{align}
where 
\begin{equation}
\kappa_{5}^{2}=\frac{\kappa^{2}}{R^{5}\Omega_{5}}\label{eq:defKappa5}
\end{equation}
and $X\left(\sigma\right)$ are the coordinates of the string world-sheet.
If we restrict to dilaton field produced by a static source then the
action takes the following form: 
\begin{align}
S_{B}\left[\Phi\right] & =\frac{\pi z_{0}R^{3}}{4\kappa_{5}^{2}}\int dtd^{2}\xbar dz\left\{ \left(\frac{1}{z}\right)^{3}\left(\left(\nabla_{i}\Phi\right)^{2}+f\left(z\right)\left(\partial_{z}\Phi\right)^{2}\right)\right\} \label{eq:dilatonActionInTermsOfz0}\\
 & +\frac{1}{4\pi l_{s}^{2}}\int d^{2}\sigma\Phi\left[X\left(\sigma\right)\right]\sqrt{g_{E}\left(X\left(\sigma\right)\right)},
\end{align}
where we have defined 
\begin{equation}
f\left(z\right)=\left(1-\frac{z^{4}}{z_{_{0}}^{4}}\right),\label{eq:def-f(z)}
\end{equation}
and have simplified the notations by using $2r_{y}=z_{0}$ and $\alpha'=l_{s}^{2}$.

\subsection{Dilaton field due to the classical open string}

The dilaton field is sourced by an open string $X\left(\sigma\right)$,
which we take to be lying in the $\left(x_{1},z\right)$ plane. 
\begin{figure}
\includegraphics[scale=0.2]{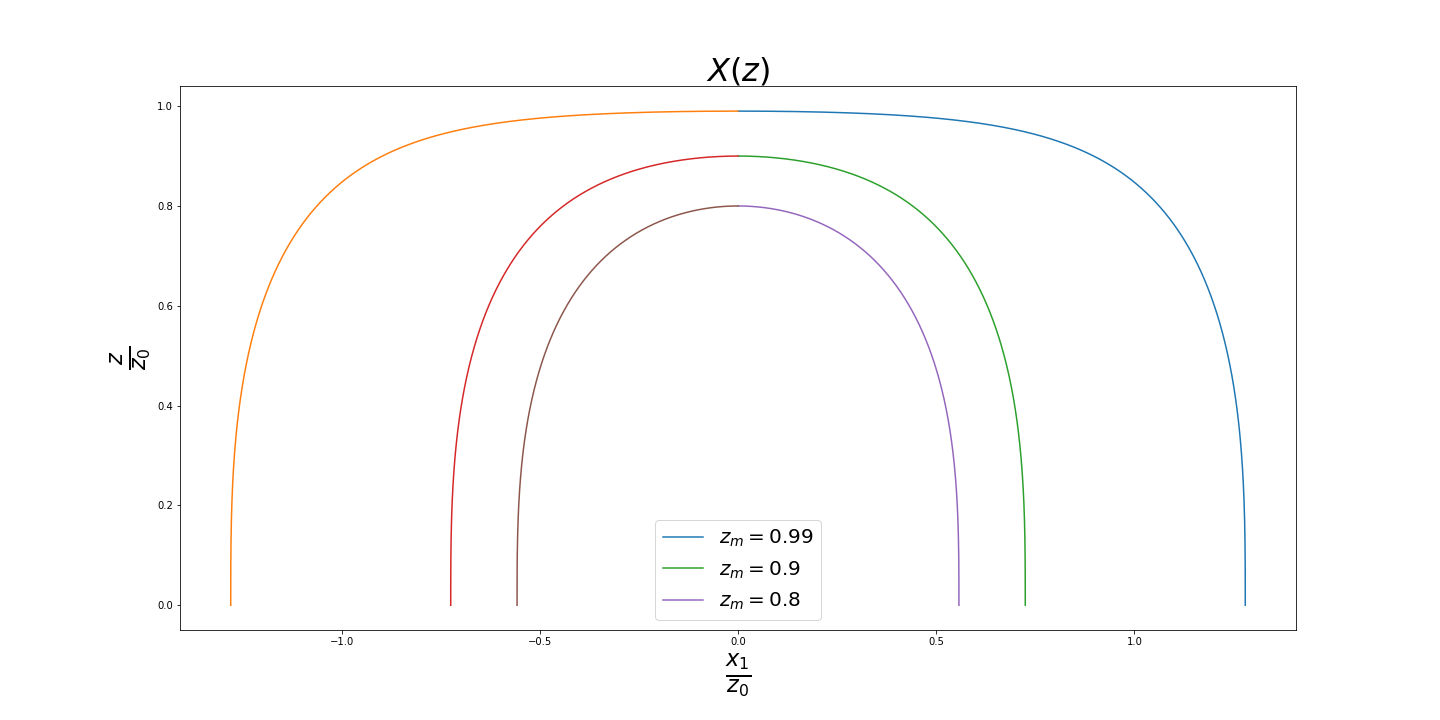}
\caption{Confining Strings for different values of $z_{m}=Z\left(0\right)$\label{fig:Confining-String. }}
\end{figure}
For numerical purposes it is convenient to parameterize the open string as 
\begin{equation}
X\left(\sigma_{1},\sigma_{2}\right)=X(t,x_{1}):=\left(t,x_{1},0,0,Z\left(x_{1}\right)\right),\label{eq:openStringParameterization}
\end{equation}
with this parameterization the action becomes 
\begin{align}
S_{B}\left[\Phi\right] & =\frac{\pi z_{0}R^{3}}{4\kappa_{5}^{2}}\int dtd^{2}\xbar dz\left\{ \left(\frac{1}{z}\right)^{3}\left(\left(\nabla_{i}\Phi\right)^{2}+f\left(z\right)\left(\partial_{z}\Phi\right)^{2}\right)\right\} \label{eq:dilatonActionInx1parameterization}\\
 & +\frac{1}{4\pi l_{s}^{2}}\int dtdx_{1}\Phi\left[X\left(x_{1}\right)\right]\sqrt{g_{E}\left(X\left(x_{1}\right)\right)}.
\end{align}
For a classical open string, using (\ref{eq:sqrt_gE}), the source
term simplifies to 
\begin{align}
S_{B}\left[\Phi\right] & =\frac{\pi z_{0}R^{3}}{4\kappa_{5}^{2}}\int dtd^{2}\xbar dz\left\{ \left(\frac{1}{z}\right)^{3}\left(\left(\nabla_{i}\Phi\right)^{2}+f\left(z\right)\left(\partial_{z}\Phi\right)^{2}\right)\right\} \label{eq:dilatonActionWithClassicalOpenString}\\
 & +\frac{z_{m}^{2}R^{2}}{4\pi l_{s}^{2}}\int dtdx_{1}\Phi\left[X\left(x_{1}\right)\right]\frac{1}{Z_{c}^{4}\left(x_{1}\right)},
\end{align}
where 
\begin{equation}
z_{m}=Z_{c}\left(0\right).\label{eq:defZm}
\end{equation}
We rescale the dilaton field by 
\begin{equation}
\bar{\Phi}=\left(\frac{\pi R^{3}}{4\kappa_{5}^{2}}\right)\Phi,\label{eq:defPhiBar}
\end{equation}
to write the action as 
\begin{align}
S_{B}\left[\Phi\right] & =\left(\frac{4\kappa_{5}^{2}}{\pi R^{3}}\right)\Biggl(z_{0}\int dtd^{2}\xbar dz\left\{ \left(\frac{1}{z}\right)^{3}\left(\left(\nabla_{i}\Phi\right)^{2}+f\left(z\right)\left(\partial_{z}\Phi\right)^{2}\right)\right\} \label{eq:scaledDilatonActionWIthOpenString}\\
+ & \frac{z_{m}^{2}\lambda^{1/2}}{4\pi}\int dtdx_{1}\Phi\left[X\left(x_{1}\right)\right]\frac{1}{Z_{c}^{4}\left(x_{1}\right)}\Biggr).
\end{align}
It is natural to measure the spatial coordinates in the units of $z_{0}$,
\begin{equation}
x_{i}\rightarrow x_{i}z_{0};\quad z\rightarrow zz_{0};\quad Z_{c}\rightarrow Z_{c}z_{0};\quad z_{m}\rightarrow z_{m}z_{0}.\label{eq:unitsOfz0}
\end{equation}
after absorbing an overall scale factor of $\left(\frac{4\kappa_{5}^{2}}{\pi R^{3}}\right)\frac{1}{z_{0}}$
we write the dilaton action as 
\begin{align}
S_{B}\left[\Phi\right] & =\int dtd^{2}\xbar dz\left\{ \left(\frac{1}{z}\right)^{3}\left(\left(\nabla_{i}\Phi\right)^{2}+f\left(z\right)\left(\partial_{z}\Phi\right)^{2}\right)\right\} \label{eq:DilatonActionInUnitsOfz0}\\
 & +\frac{z_{m}^{2}\lambda^{1/2}}{4\pi}\int dtdx_{1}\Phi\left[X\left(x_{1}\right)\right]\frac{1}{Z_{c}^{4}\left(x_{1}\right)}.
\end{align}
It will also be useful to write the bulk action as, 
\begin{equation}
S_{B}\left[\Phi\right]=\int dtd^{2}\xbar dz\left\{ \left(\frac{1}{z}\right)^{3}\left(\left(\nabla_{i}\Phi\right)^{2}+f\left(z\right)\left(\partial_{z}\Phi\right)^{2}\right)+\Phi J\right\} ,\label{eq:bulkActionWithCurrent}
\end{equation}
where 
\begin{equation}
J=\frac{z_{m}^{2}\lambda^{1/2}}{4\pi}\delta(x_{2})\delta\left(z-Z_{c}\left(x_{1}\right)\right)\frac{1}{z^{4}}.\label{eq:defJ}
\end{equation}

\subsection{Classical flux-tube in $\Qcdthree$}

The partition function for $\Qcdthree$ in the presence of a $q\bar{q}$
pair and a source term for $\Ofsquare$ can be schematically written
as 
\begin{equation}
Z_{\Qcdthree}[\phi]=\int[dA]\exp\left\{ -S_{\Qcdthree}^{q\bar{q}}[A]+2\pi r_{y}\int dtd^{2}\xbar\mathcal{O}_{F^{2}}\left(t,\bar{x}\right)\phi\left(t,\xbar\right)\right\} .\label{eq:ZqcdThree}
\end{equation}
According to the dictionary for the gauge/gravity duality the source
term $\phi\left(\xbar\right)$ is related to the asymptotic value
of the dilaton field, again after the Kaluza-Klein reduction over
the compact $y$ direction, 
\begin{align}
\phi\left(t,\xbar\right) & =\lim_{z\rightarrow0}\Phi_{c}\left(z,t,\xbar\right)=\lim_{z\rightarrow0}\left(\Phi_{H}\left(z,t,\xbar\right)+\Phi_{s}\left(z,t,\xbar\right)\right),\label{eq:sourceInTermsOfDilatonField}\\
\lim_{z\rightarrow0} & \Phi_{H}\left(z,t,\bar{x}\right)=\phi\left(t,\xbar\right),\label{eq:defPhiHomogenous}\\
 & \lim_{z\rightarrow0}\Phi_{s}\left(z,t,\xbar\right)=0,\label{eq:boundaryConditonOnSourcedDilatonField}
\end{align}
where $\Phi_{H}\left(z,t,\xbar\right)$ is the solution of the homogenous
dilaton field equation with the boundary condition given by (\ref{eq:defPhiHomogenous}),
while $\Phi_{s}\left(z,\xbar\right)$ is the classical dilaton field
produced by the fundamental open string in the bulk. $\Phi_{H}$ is
the so called non-normalizable mode while $\Phi_{s}$ is the normalizable
mode. It will be convenient to write the source term in terms of the
rescaled bulk field (\ref{eq:defPhiBar}) and to measure the spatial
coordinates $\xbar$ at the boundary in the units of $z_{0}=1$ (\ref{eq:unitsOfz0}),
but we will continue to measure $\Ofsquare$ in normal units in which
$z_{0}\ne1$, 
\begin{equation}
Z_{\Qcdthree}\left[\phi\right]=\int[dA]\exp\left\{ -S_{\Qcdthree}[A]+\pi z_{0}\left(\frac{4\kappa_{5}^{2}}{\pi R^{3}}\right)z_{0}^{2}\int dtd^{2}\xbar\mathcal{O}_{F^{2}}\left(t,\xbar\right)\phi\left(t,\xbar\right)\right\} .\label{eq:ZQcdThreeRescaled}
\end{equation}

Using the dictionary provided by the gauge/gravity duality in the
limit $N\rightarrow\infty$ and $\lambda>>1$ 
\begin{equation}
-\frac{1}{z_{0}^{4}}\left(\frac{\delta\bar{S}}{\delta\Phi_{s}}\right)_{z=0}=\,<\Omega|\Ofsquare\left(t,\xbar\right)|\Omega>_{q\bar{q}}.\label{eq:expectationvalueOfFsquareUsingGG}
\end{equation}
Evaluating the variation about the classical solution we find that
the only non-vanishing contribution comes from 
\begin{equation}
\delta S\left[\Phi_{c}\right]=\int d^{3}x\left[\frac{2f\left(z\right)}{z^{3}}\left(\partial_{z}\Phi_{s}\right)\delta\Phi\right]_{z=z_{0}}^{z=0}.\label{eq:zBoundaryContributionToDeltaS}
\end{equation}
Since the bulk ends smoothly at $z_{0}$, we have the following boundary
condition at $z_{0}$ 
\begin{equation}
\left(\frac{\partial\Phi}{\partial z}\right)_{z=z_{0}}=0,\label{eq:boundaryConditonAtz0}
\end{equation}
we obtain 
\begin{equation}
<\Omega|\Ofsquare\left(t,\xbar\right)|\Omega>_{q\bar{q}}^{X\left(\sigma\right)}=\lim_{a\rightarrow0}\left(-\frac{1}{z_{0}^{4}}\frac{2f\left(a\right)}{a^{3}}\left(\partial_{z}\Phi_{s}\right)_{z=a}\right).\label{eq:ExpectationValueOfTrFsquare}
\end{equation}
where the superscript $X\left(\sigma\right)$ on the left hand side
is a reminder that this is the expectation value due to a particular
string configuration, $X\left(\sigma\right)$, in our case the classical
string configuration.

\subsection{Prescription for calculating flux-profile in $\Qcdthree$}

Up till now we have considered a classical open string joining the
quark and the antiquark as the source of the dilaton field, which
in turn via (\ref{eq:ExpectationValueOfTrFsquare}), describes a corresponding
flux-tube in the boundary theory. We will refer to the flux-tube induced
by the classical open string configuration as the classical flux-tube.
As was noted in the introduction, the flux-profile in the boundary
theory can be thought of as a superposition of flux-tubes induced
by the various string configurations making up the open string wave-function.
We can formally incorporate the fluctuations of the open string by
writing the partition function for the dilaton field as

\begin{align}
Z_{Gravity}\left[\phi\left(\bar{x}\right)\right] & =\int\left[dX\right]\exp\left\{ -S_{NG}\left[X\right]\right\} \int\left[d\Phi\right]\exp\left\{ -S_{B}[\Phi;X]\right\} ,\label{eq:ZgravityWithStringFluctuations}\\
 & =\int\left[dX\right]\exp\left\{ -S_{NG}\left[X\right]\right\} \exp\left\{ -S_{B}\left[\Phi_{c}\left[X\right]\right]\right\} ,
\end{align}
where $X\left(\sigma\right)$ are the open string world sheets corresponding
to strings that start at a quark and end at an antiquark, $\Phi$
represent the dilaton configurations that satisfy the boundary conditions
given by (\ref{eq:sourceInTermsOfDilatonField},\ref{eq:defPhiHomogenous},\ref{eq:boundaryConditonOnSourcedDilatonField}),
$S_{NG}$ is the Nambu-Goto action for an open string (\ref{eq:NGinStringFrame})
and $S_{B}\left[\Phi;X\right]$ is the dilaton action in the presence
of an open string (\ref{eq:DilatonActionInUnitsOfz0}). The functional
integral over the dilaton field has been approximated by its classical
value for $\lambda>>1$ and $N\rightarrow\infty$. Formally the flux-profile
in the boundary theory can be written as 
\begin{eqnarray}
<\Omega|\Ofsquare\left(t,\xbar\right)|\Omega>_{q\bar{q}} & = & \int\frac{\left[dX\right]}{Z_{NG}}\exp\left\{ -S_{NG}\left[X\right]\right\} \left(-\frac{1}{z_{0}^{4}}\left(\frac{\delta\bar{S}}{\delta\Phi[X]}\right)_{\Phi_{c}\left(z=0,\xbar\right)}\right)\nonumber \\
 & = & \int\frac{\left[dX\right]}{Z_{NG}}\exp\left\{ -S_{NG}\left[X\right]\right\} <\Omega|\Ofsquare\left(t,\xbar\right)|\Omega>_{q\bar{q}}^{X\left(\sigma\right)}\label{eq:fluxProfileFromFlux-Tubes}
\end{eqnarray}

where 
\begin{equation}
Z_{NG}=\int\left[dX\right]\exp\left\{ -S_{NG}\left[X\right]\right\} ,\label{eq:ZnambuGoto}
\end{equation}
is the partition function of the Nambu-Goto string.

We will not make an attempt to evaluate the above functional integral
but will try and get a qualitatively understanding by evaluating the
shape of the flux-tubes for a class of fluctuating open strings.

\section{Framework for numerical calculations\label{sec:Framework-for-numerical}}

To calculate the profile of the flux-tube using (\ref{eq:ExpectationValueOfTrFsquare})
we need to solve for the classical dilaton field, $\Phi_{c}\left(x,z\right)$,
sourced by an open string. We will solve for the classical dilaton
field numerically, to do so we need a discrete version of the dilaton
field equation. A convenient way of doing that is by first discretizing
the dilaton action \cite{RevModPhys.56.1,koonin1998computational}.

\subsection{Discretized dilaton action}

Since we are working with static fields therefore it is more convenient
to discretize 
\begin{align}
L\left[\Phi\right] & =\int d^{2}\xbar dz\left\{ \left(\frac{1}{z}\right)^{3}\left(\left(\nabla_{i}\Phi\right)^{2}+f\left(z\right)\left(\partial_{z}\Phi\right)^{2}\right)\right\} \label{eq:defL}\\
 & +\frac{z_{m}^{2}\lambda^{1/2}}{4\pi}\int dx_{1}\Phi\left[X\left(x_{1}\right)\right]\frac{1}{Z_{c}^{4}\left(x_{1}\right)}.
\end{align}
Consider first the ``kinetic'' part of the action 
\begin{equation}
L_{1}\left[\Phi\right]=\int_{-\infty}^{\infty}d^{2}x\int_{0}^{1}dz\left\{ A\left(z\right)\left(\frac{\partial\Phi}{\partial x_{1}}\right)^{2}+A\left(z\right)\left(\frac{\partial\Phi}{\partial x_{2}}\right)^{2}+B\left(z\right)\left(\partial_{z}\Phi\right)^{2}\right\} ,\label{eq:LPhi}
\end{equation}
where we have defined 
\begin{equation}
A\left(z\right)=\frac{1}{z^{3}};\quad B\left(z\right)=\frac{f\left(z\right)}{z^{3}}=\frac{\left(1-z^{4}\right)}{z^{3}}.\label{eq:defAandB}
\end{equation}
We put the system in a three-dimensional box: 
\begin{align}
-\frac{L_{1}}{2} & \le x_{1}\le\frac{L_{1}}{2},\nonumber \\
-\frac{L_{2}}{2} & \le x_{2}\le\frac{L_{2}}{2},\nonumber \\
0 & \le z\le1,\label{eq:sizeOftheBox}
\end{align}
$L_{i}$ acting as an infrared cut-off. Next we discretize the spatial
coordinates by introducing an anisotropic lattice 
\begin{align}
x_{i} & =n_{i}b,\quad n_{i}\in\left(-N_{i},N_{i}\right),\quad L_{i}=2N_{i}b,\nonumber \\
z & =n_{3}a,\quad n_{3}\in\left(0,N_{3}\right),\quad1=N_{3}a,\label{eq:Lattice}\\
b & =\sigma a,\quad\sigma>0.\label{eq:defSigma}
\end{align}
Having an anisotropic lattice will be convenient for numerical solution
of the dilaton field equation. We denote the value of the dilaton
field on the points of the lattice as 
\begin{equation}
\Phi\left(x_{1},x_{2},x_{3}\right)=\Phi\left(n_{1}b,n_{2}b,n_{3}a\right)=\Phi\left(n_{1},n_{2},n_{3}\right).\label{eq:PhiOnLattice}
\end{equation}
Now we can discretize the action (\ref{eq:LPhi}) by calculating the
integrand at the centre of the elementary cubes 
\begin{align}
L_{1}^{\text{lattice}}\left[\Phi\right] & =\sum_{\left(n_{1,}n_{2},n_{3}\right)}b^{2}a\left\{ A\left(n_{3}+1/2\right)\left(\frac{\Phi\left(n_{1}+1,n_{2},n_{3}\right)-\Phi\left(n_{1},n_{2},n_{3}\right)}{b}\right)^{2}\right.\nonumber \\
 & +A\left(n_{3}+1/2\right)\left(\frac{\Phi\left(n_{1},n_{2}+1,n_{3}\right)-\Phi\left(n_{1},n_{2},n_{3}\right)}{b}\right)^{2}\nonumber \\
 & \left.+B\left(n_{3}+1/2\right)\left(\frac{\Phi\left(n_{1},n_{2},n_{3}+1\right)-\Phi\left(n_{1},n_{2},n_{3}\right)}{a}\right)^{2}\right\} .\label{eq:L1onLattice}
\end{align}
In terms of the scale factor $\sigma$, given by (\ref{eq:defSigma}),
the action becomes 
\begin{align}
L_{1}^{\text{lattice}}\left[\Phi\right] & =\sum_{\left(n_{1,}n_{2},n_{3}\right)}a\left\{ A\left(n_{3}+1/2\right)\left(\Phi\left(n_{1}+1,n_{2},n_{3}\right)-\Phi\left(n_{1},n_{2},n_{3}\right)\right)^{2}\right.\nonumber \\
 & +A\left(n_{3}+1/2\right)\left(\Phi\left(n_{1},n_{2}+1,n_{3}\right)-\Phi\left(n_{1},n_{2},n_{3}\right)\right)^{2}\nonumber \\
 & \left.+\sigma^{2}B\left(n_{3}+1/2\right)\left(\Phi\left(n_{1},n_{2},n_{3}+1\right)-\Phi\left(n_{1},n_{2},n_{3}\right)\right)^{2}\right\} .\label{eq:L1onLattice-1}
\end{align}

The source part of the action (\ref{eq:defL}) for the classical open
string is 
\begin{equation}
L_{s}=\frac{z_{m}^{2}\lambda^{1/2}}{4\pi}\int dx_{1}\Phi\left[X\left(x_{1}\right)\right]\frac{1}{Z_{c}^{4}\left(x_{1}\right)},\label{eq:LsourceOpenString}
\end{equation}
and can be discretized as 
\begin{equation}
L_{s}^{\text{lattice}}[\Phi]=\frac{z_{m}^{2}\lambda^{1/2}}{4\pi}\sum_{n_{1}}\sigma a\left\{ \Phi\left(n_{1},0,Z_{c}\left(n_{1}\right)\right)\frac{1}{Z_{c}^{4}\left(n_{1}\right)}\right\} .\label{eq:LsourceDiscrete}
\end{equation}

\subsection{The discrete field equation}

We obtain the discrete field equation by minimizing the discrete Lagrangian
\begin{equation}
\frac{\partial}{\partial\Phi\left(m_{1},m_{2},m_{3}\right)}\left\{ L_{1}^{\text{lattice}}+L_{s}^{\text{lattice}}\right\} =0,\label{eq:discretePrincipleOfLeastAction}
\end{equation}
which leads to the following discrete field equation for the dilaton
field 
\begin{align}
\left(4A\left(m_{3}+1/2\right)+\sigma^{2}B\left(m_{3}-1/2\right)+\sigma^{2}B\left(m_{3}+1/2\right)\right)\Phi\left(m_{1},m_{2},m_{3}\right)\nonumber \\
-A\left(m_{3}+1/2\right)\left(\Phi\left(m_{1}-1,m_{2},m_{3}\right)+\Phi\left(m_{1}+1,m_{2},m_{3}\right)\right)\nonumber \\
-A\left(m_{3}+1/2\right)\left(\Phi\left(m_{1},m_{2}-1,m_{3}\right)+\Phi\left(m_{1},m_{2}+1,m_{3}\right)\right)\nonumber \\
-\left(\sigma^{2}B\left(m_{3}-1/2\right)\Phi\left(m_{1},m_{2},m_{3}-1\right)+\sigma^{2}B\left(m_{3}+1/2\right)\Phi\left(m_{1},m_{2},m_{3}+1\right)\right)\nonumber \\
+\frac{\sigma z_{m}^{2}\lambda^{1/2}}{8\pi}\left(\delta\left(0,m_{2}\right)\delta\left(Z_{c}\left(m_{1}\right),m_{3}\right)\frac{1}{Z_{c}^{4}\left(m_{1}\right)}\right) & =0\label{eq:discreteDilatonFieldEquation}
\end{align}

The dilaton field equation on the lattice, (\ref{eq:discreteDilatonFieldEquation}),
has to be supplemented by the boundary conditions on our box (\ref{eq:sizeOftheBox},\ref{eq:Lattice}).
We are interested only in the dilaton field which is sourced by the
open string connecting a quark and an anti-quark, for such field we
expect 
\begin{equation}
\lim_{|\xbar|\rightarrow\infty}\Phi\left(\xbar,z\right)=0.\label{eq:boundarConditionsOnXbar}
\end{equation}
On our lattice (\ref{eq:Lattice}) this translates into 
\begin{equation}
\Phi\left(\pm N_{1},n_{2},n_{3}\right)=0,\quad\Phi\left(n_{1},\pm N_{2},n_{3}\right)=0.\label{eq:latticeBConXbar}
\end{equation}
The bulk smoothly ends at $z_{0}$ that results in 
\[
\left(\frac{\partial\Phi}{\partial z}\right)_{z=z_{0}}=0,
\]
which translates on our lattice as 
\begin{equation}
\Phi\left(\bar{n},N_{3}\right)=\Phi\left(\bar{n},N_{3}-1\right).\label{eq:latticeBCatZ0}
\end{equation}
Next we have to impose the boundary condition at the conformal infinity
$z=0$ (\ref{eq:boundaryConditonOnSourcedDilatonField}) which translates
on the lattice as 
\begin{equation}
\Phi\left(\bar{n},0\right)=0.\label{eq:latticeBCatZ=00003D00003D0}
\end{equation}

\subsection{Numerical definition of the flux-tube}

With all these preliminaries in place, we can translate the formula
(\ref{eq:ExpectationValueOfTrFsquare}) on to the lattice 
\begin{equation}
<\Omega|\Ofsquare\left(\bar{n}\right)|\Omega>_{q\bar{q}}^{X\left(\sigma\right)}=-\frac{1}{z_{0}^{4}}B\left(1\right)\left(\frac{\Phi\left(\bar{n},1\right)-\Phi\left(\bar{n},0\right)}{a}\right),\label{eq:latticeExpValueOfTrFsquare}
\end{equation}
further using the boundary condition (\ref{eq:latticeBCatZ=00003D00003D0})
we get 
\begin{equation}
<\Omega|\Ofsquare\left(\bar{n};\right)|\Omega>_{q\bar{q}}^{X\left(\sigma\right)}=-\frac{1}{z_{0}^{4}}B\left(1\right)\left(\frac{\Phi\left(\bar{n},1\right)}{a}\right).\label{eq:latticeExpFor<O>OnLattice}
\end{equation}

\section{Shapes of confining flux-tubes\label{sec:Shape-of-the-1}}

Now we are ready to numerically explore the formation of the flux-tubes
in our theory. We will consider the flux-tube corresponding to the classical open
string configuration, this is the flux-tube which leads to the linear confining potential for large separation of quark and anti-quark. In solving (\ref{eq:discreteDilatonFieldEquation}
) the string-dilaton coupling 
\begin{equation}
g_{sd}=\frac{\lambda^{1/2}}{8\pi}\label{eq:defDilatonStringCouplingConstant}
\end{equation}
appears only as a multiplying constant in the source term, therefore
without any loss of generality we can solve the equation for an arbitrary
value of $g_{sd}$. The shape of the flux-tube will be independent
of the value of $g_{sd}$, the actual value of the action density
will depend on $g_{sd}$ but knowing its value for a given $g_{sd}$
we can easily obtain the value for any other $g_{sd}$ by simple rescaling.
In numerically solving (\ref{eq:discreteDilatonFieldEquation}) to
desired precision, one faces certain number of difficulties which
we point out in the appendix (\ref{sec:Appendix:-Numerical-solution})
and there we also describe the strategies we have used to overcome
them.

\subsection{Emergence of a classical flux-tube\label{subsec:Emergence-of-fluxtube}}

\begin{figure}
\includegraphics[scale=0.32]{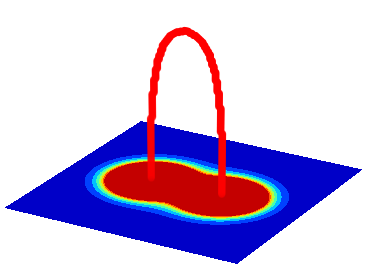}\includegraphics[scale=0.32]{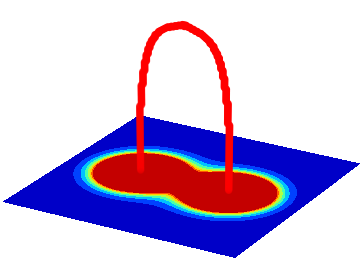}\includegraphics[scale=0.2]{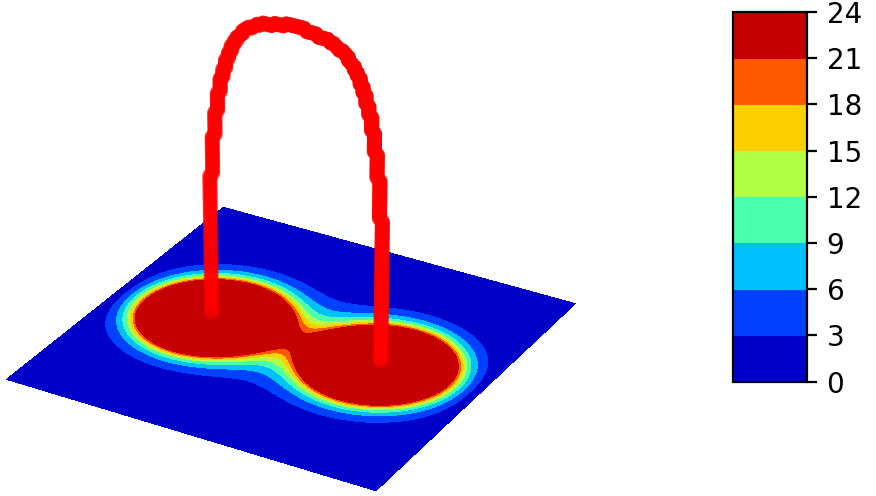}

\caption{Normalized profile of a flux-tube with the corresponding classical
open string in the bulk for $L_{q\bar{q}}=1.56\text{,\,1.95\text{ and}\,2.35}z_{0}$.\label{fig:Emerg-FLux-Tube}}
\end{figure}

We will solve the dilaton equation (\ref{eq:discreteDilatonFieldEquation})
on an anisotropic lattice with the lattice constant along the bulk
direction $z$ equal to 
\begin{equation}
a=\frac{1.0}{128}z_{0}\label{eq:aFor128cube}
\end{equation}
and with the lattice constant along the boundary directions $(x_{1},x_{2})$
\begin{equation}
b=5a=0.0390z_{0}.\label{eq:bFor128cube}
\end{equation}
The size of the resulting lattice is 
\begin{equation}
V_{lattice}=128a\times128b\times128b=z_{0}\times5z_{0}\times5_{z0}\label{eq:latticeSizeIn3dim}
\end{equation}
With such a lattice we can explore a maximum quark - antiquark distance
$L_{q\bar{q}}=2.34z_{0}$, beyond that the boundary of the lattices
start distorting the profile. For each value of $L_{q\bar{q}}$ the
equation (\ref{eq:discreteDilatonFieldEquation}) was solved using
the full multi-grid algorithm till the norm of the residue (\ref{eq:normResidue-3dim})
was of the order $10^{-2}$ (see the accompanying python notebook,
which were used to obtain the numerical results, for the details.)
In delineating the shape of a flux-tube one faces the difficulty that
the value of the action density diverges\footnote{This divergence is related to the bare mass in the boundary theory, see for e.g. \cite{Kol:2010fq} for a review. } at the position of the quark
and the anti-quark which completely masks the shape of the flux-tube.
Since our interest is just to delineate and display the shape of the
flux-tube we will normalize the flux-tube profile, 
\begin{equation}
P^{X\left(\sigma\right)}\left(x_{1},x_{2}\right)=<\Omega|\Ofsquare\left(x_{1},x_{2}\right)|\Omega>_{q\bar{q}}^{X\left(\sigma\right)}\label{eq:fluxTubeProfile}
\end{equation}
in the following manner: 
\begin{equation}
\text{If }P^{X\left(\sigma\right)}\left(x_{1},x_{2}\right)>P^{X\left(\sigma\right)}(x_{Lc},x_{Tc})\text{ then }P^{X\left(\sigma\right)}(x_{1},x_{2})=P^{X\left(\sigma\right)}(x_{Lc},x_{Tc}),\label{eq:FluxNormalization-1}
\end{equation}
where $P^{X\left(\sigma\right)}(x_{Lc},x_{Tc})$ is the value of the
flux-tube profile at the central point along the line joining the
quark and the anti-quark when $L_{q\bar{q}}=2.34z_{0}$
We present the results in Figs.(\ref{fig:Emerg-FLux-Tube}). 
It is important to note that our results for these flux-tube profiles
are only qualitative in nature as the inter quark distance is comparable
to the size $r_{y}$ of the compact $y$ direction and we are averaging
dilaton field over the $y$ direction. Even with this smearing of
profile what we can glean is that a flux-tube starts appearing when
the inter-quark distance is of the order of $3z_{0}$.

\subsection{Approximate shape of a long but finite classical flux-tube\label{subsec:Rectangular-Fluxtube}}

\begin{figure}
\includegraphics[scale=0.75]{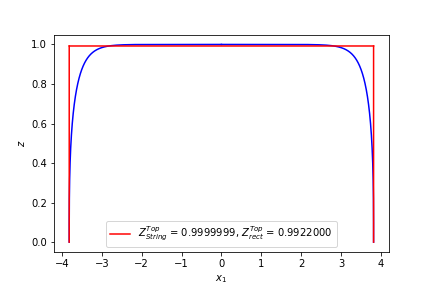}

\caption{Open string corresponding to a long flux-tube and its approximation
by a rectangular source of the dilaton field.\label{fig:Open-string-Rectange}}
\end{figure}

As one increases the distance between a quark and an antiquark the
corresponding open string moves closer and closer to $z_{0}$, asymptotically
approaching $z_{0}$. To represent such an open string on a lattice
requires that the lattice constant in the $z$ direction should tend
to zero or equivalently the number of points along the $z$ direction
tend to infinity, making the problem computationally prohibitive.
We circumvent the problem approximately by replacing the open string
by a rectangular source of dilaton whose top is placed at 
\begin{equation}
z_{Rect}^{Top}=z_{0}-a,\label{eq:defZtop}
\end{equation}
where $a$ is the lattice constant along the $z$ direction (see Fig.(\ref{fig:Open-string-Rectange})) 

\begin{figure}
\includegraphics[scale=0.4]{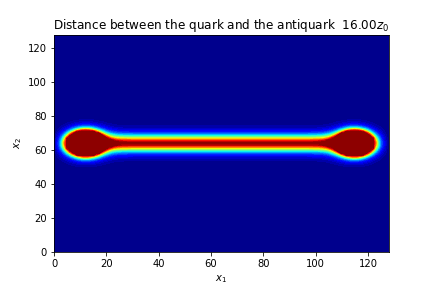}\includegraphics[scale=0.4]{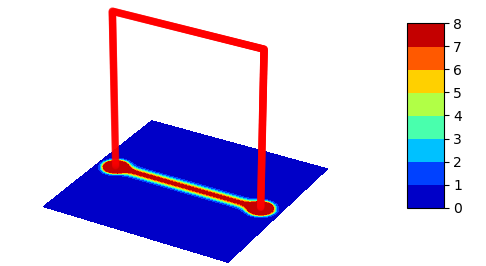}

\caption{Long flux-tube created by a rectangular source of dilaton field\label{fig:Long-flux-tube-created}}
\end{figure}
\begin{figure}
\includegraphics[scale=0.45]{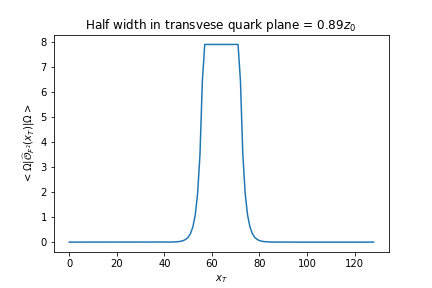}\includegraphics[scale=0.45]{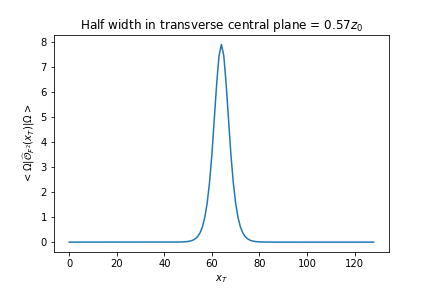}

\caption{Thickness of a long classical flux-tube\label{fig:ThicknessRectFluxTube}}
\end{figure}

In Fig.(\ref{fig:Long-flux-tube-created}) and Fig.(\ref{fig:ThicknessRectFluxTube})
we exhibit the flux-tube obtained using a rectangular dilaton source
on an anisotropic lattice with the lattice constant $a$ along $z$
direction is 
\begin{equation}
a=\frac{1.0}{128}z_{0}\label{eq:aFor256Cube}
\end{equation}
and with lattice constant $b$ along the boundary directions $(x_{1},x_{2})$
\begin{equation}
b=20.0a.\label{eq:bFor256Cube}
\end{equation}
The source term for rectangular open string configuration can be easily
obtained using (\ref{eq:DilatonActionInUnitsOfz0}) and is noted in
(\ref{subsec:Dilaton-Rect-Source}). As in the previous case, the
dilaton field equation was solved using the full multi-grid algorithm
till the norm of the residue (\ref{eq:normResidue-3dim}) was of the
order $10^{-2}$. With in the approximations that we have made, one
can clearly see formation of a flux-tube with a finite intrinsic thickness.

\subsection{Intrinsic thickness of an infinitely long classical flux-tube \label{sec:Intrinsic-Thicknes-of}}

We will next consider the situation where the quark and the anti-quark
are separated by a distance much greater than $r_{y}$, or formally
when 
\begin{equation}
L_{q\bar{q}}\rightarrow\infty.\label{eq:defLongFluxTube}
\end{equation}
Gravitationally such a situation is described by an infinitely long
open string placed at $z=z_{0}$. An approximate analytic solution
for this case was already obtained in \cite{Danielsson:1998wt}.
On a lattice the profile of infinitely long open string is given by
\begin{equation}
Z_{c}^{\text{lattice}}\left(m_{1}\right)=Z_{c}\left(0\right)=z_{m}=aN_{m}.\label{eq:veryLongClassicalOpenString}
\end{equation}
The dilaton field sourced by such a long open string depends only
on the transverse coordinate $x_{2}=am_{2}$ and on the radial coordinate
$z=am_{3}$, 
\begin{align}
\left(2A\left(m_{3}+1/2\right)+B\left(m_{3}-1/2\right)+B\left(m_{3}+1/2\right)\right)\Phi\left(m_{2},m_{3}\right)\nonumber \\
-A\left(m_{3}+1/2\right)\left(\Phi\left(m_{2}-1,m_{3}\right)+\Phi\left(m_{2}+1,m_{3}\right)\right)\nonumber \\
-\left(B\left(m_{3}-1/2\right)\Phi\left(m_{2},m_{3}-1\right)+B\left(m_{3}+1/2\right)\Phi\left(m_{2},m_{3}+1\right)\right)\nonumber \\
+\left(\frac{\lambda^{1/2}}{8\pi}\right)\frac{\delta\left(0,m_{2}\right)\delta\left(Z_{c}\left(0\right),m_{3}\right)}{Z_{c}^{2}\left(0\right)} & =0.\label{eq:twoDimDiscreteDilatonFieldEquation}
\end{align}
We will solve the above equation numerically and use (\ref{eq:latticeExpFor<O>OnLattice})
to plot the intrinsic shape of the flux tube.

\subsection{Numerical Results: Shape of a very long classical flux-tube\label{subsec:Numerical-Infinite-String}}

\begin{figure}
\includegraphics[scale=0.4]{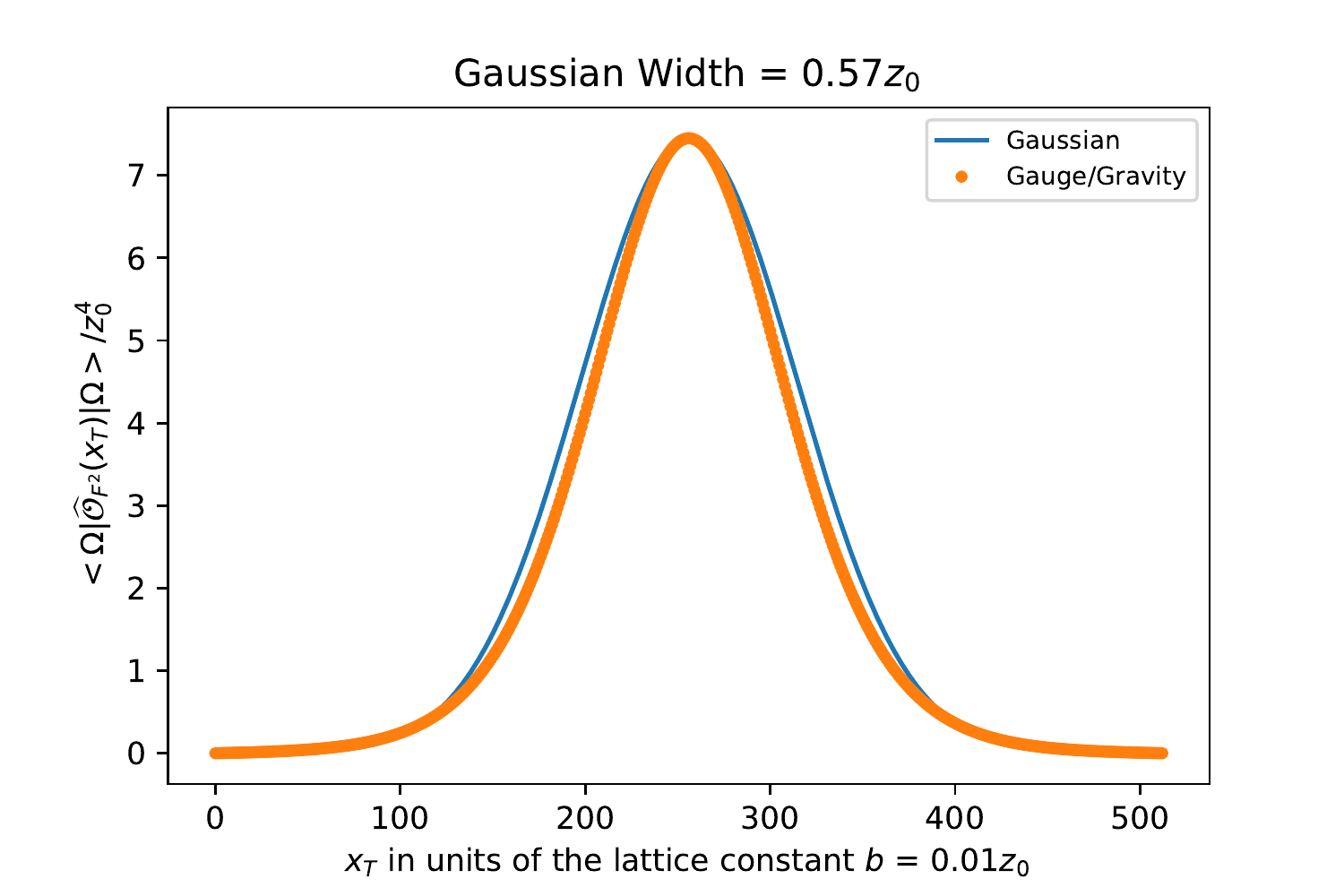}\includegraphics[scale=0.4]{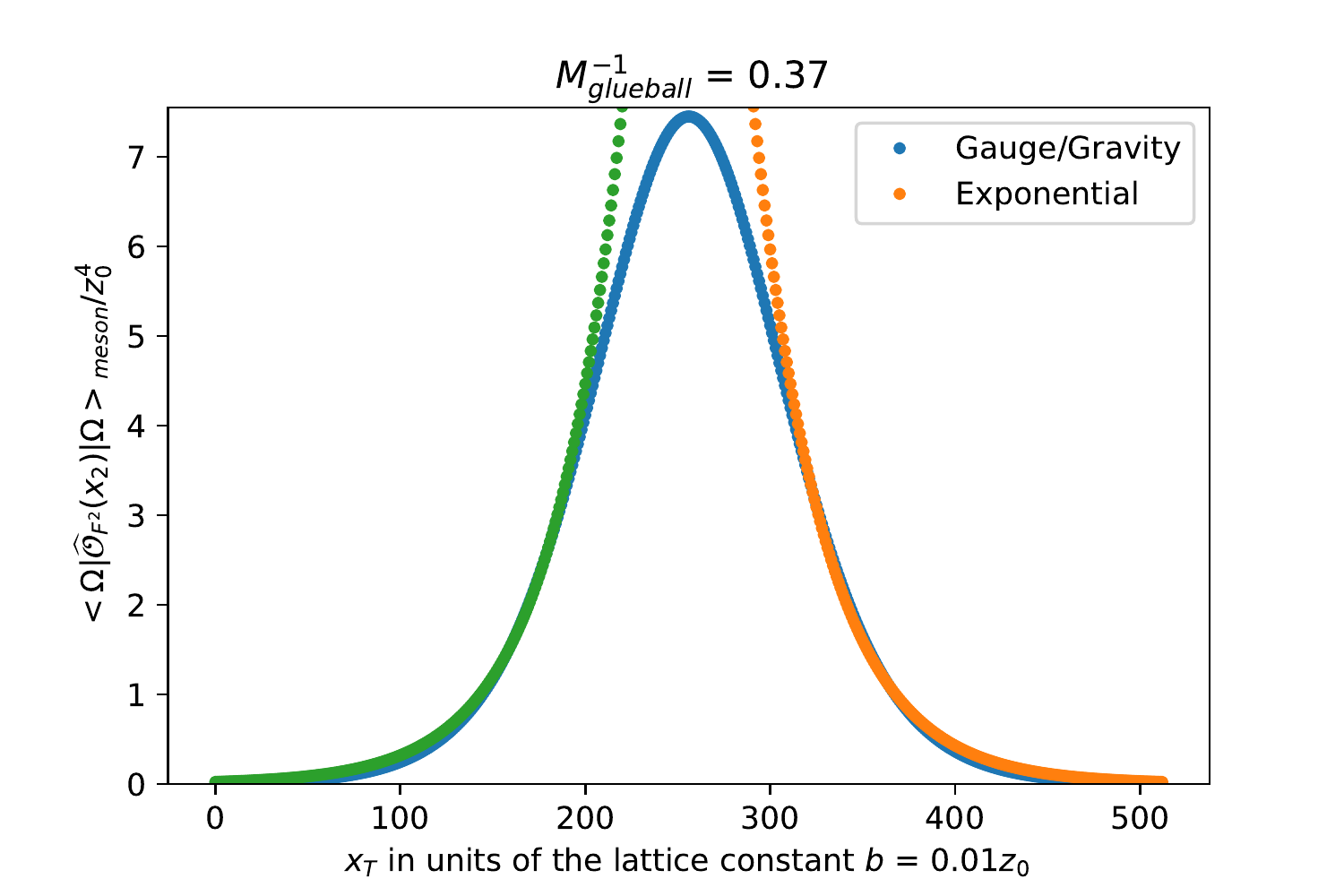}\caption{Intrinsic shape of a infinitely long confining flux-tube and its fit
to an exponential and Gaussian function\label{fig:Intrinsic-shape-details}}
\end{figure}

The transverse profile of a long classical flux-tube, $L_{q\bar{q}}\rightarrow\infty$,
can be defined as 
\begin{equation}
P_{T}^{X_{c}\left(\sigma\right)}\left(x_{T}\right)=\lim_{L_{q\bar{q}}\rightarrow\infty}<\Omega|\Ofsquare\left(x_{1}=0;x_{2}=x_{T}\right)|\Omega>_{q\bar{q}}^{X_{c}\left(\sigma\right)}.\label{eq:TransverseProfile-1}
\end{equation}
If the flux-tube is formed because of the finite correlation-length
of flux-lines, as suggested in \cite{Wilson:1977nj}, then we expect
that the transverse profile should take the following form 
\begin{equation}
\lim_{|x_{T}|\rightarrow\infty}P_{T}\left(x_{T}\right)=C_{0}\exp\left(-\frac{|x_{T}|}{\xi_{FT}}\right)=C_{0}\exp\left(-|x_{T}|M_{\text{glueball}}^{-1}\right),),\label{eq:ProfileForConfiningFluxTube-1}
\end{equation}
where $C_{0}$ is a constant of mass dimension four and $\xi_{FT}$ 
is the correlation length of the flux-lines (see the discussion in
section 2 of \cite{Wilson:1977nj}) and
is a measure of the intrinsic thickness of the flux-tube. The inverse of  $\xi_{FT}$ is, on dimensional grounds, related to the lightest glueball mass.
\begin{figure}
\includegraphics[scale=0.4]{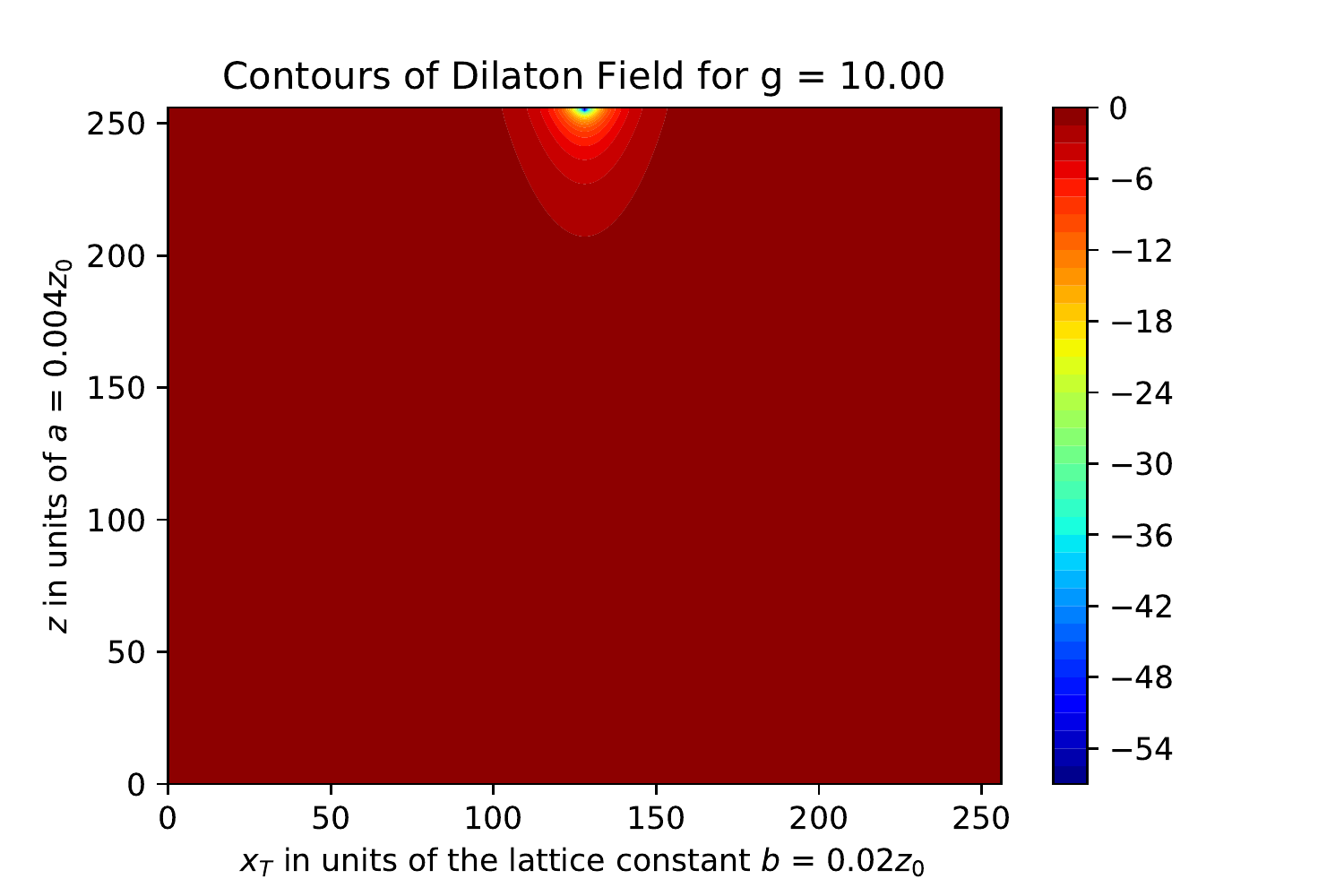}\includegraphics[scale=0.4]{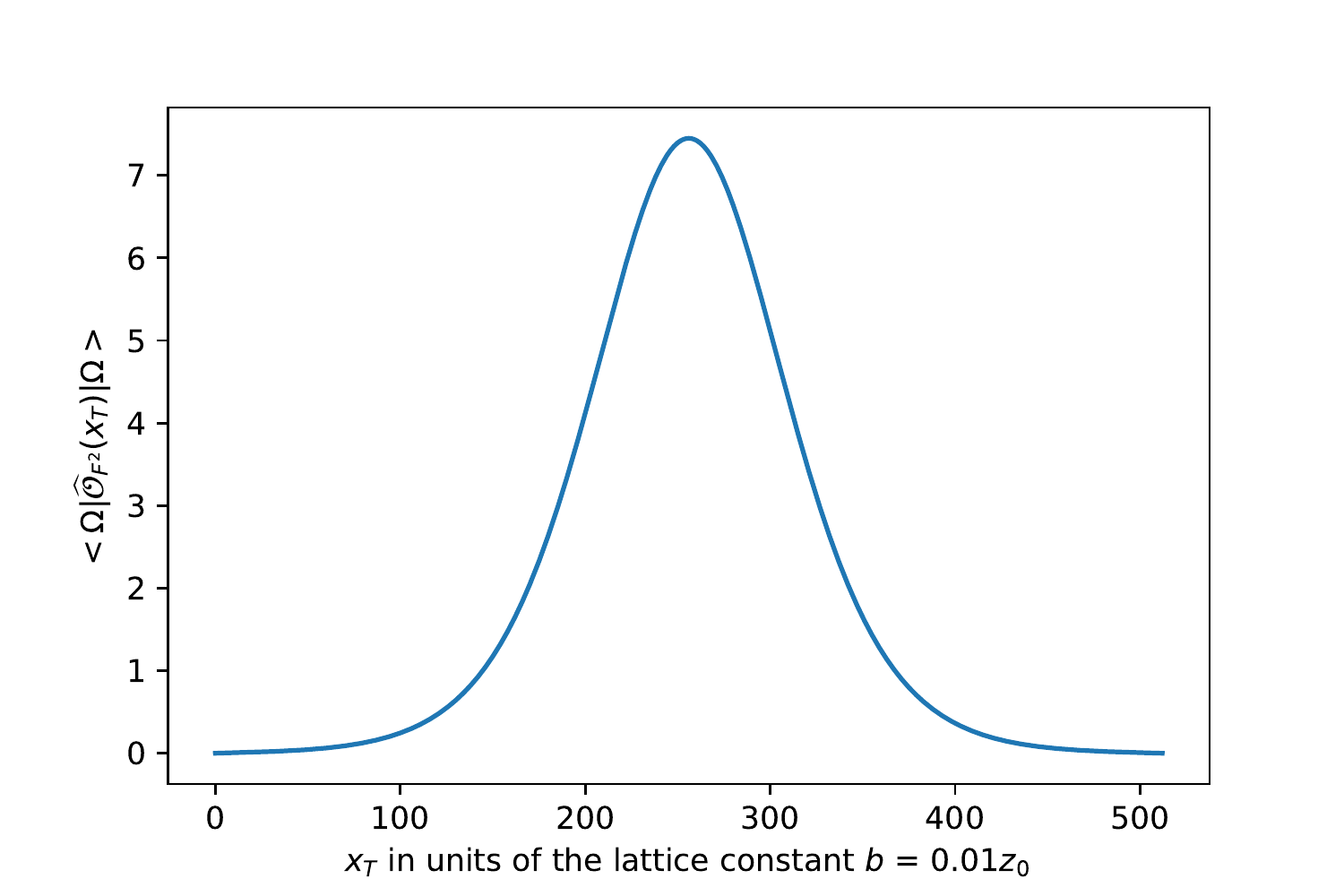}\caption{Dilaton field due to a infinitely long open string along the $x_{1}$
axis and the resulting confining flux-tube on the boundary.\label{fig:Dilaton-field-due}}
\end{figure}
The results of numerical calculations are presented in Fig.(\ref{fig:Dilaton-field-due})
and more detailed description of fitting of the data with a Gaussian
and an exponential function is provided in Fig.(\ref{fig:Intrinsic-shape-details}).
Because of the two dimensional nature of the problem the dilaton field
equation could be solved with great precision, norm of the residue
(\ref{eq:normResidue-2dim}) being of the order of $10^{-6}$.

\begin{table}[h]
\centering %
\begin{tabular}{cc}
\hline 
$L_{q\bar{q}}$  & Half width along the line $x_{1}=L_{q\bar{q}}/2$ \tabularnewline
\hline 
1.56  & 0.65 \tabularnewline
\hline 
1.95  & 0.61 \tabularnewline
\hline 
2.34  & 0.58 \tabularnewline
\hline 
$16.00\text{ rect. string}$  & 0.57 \tabularnewline
\hline 
$\infty$  & 0.57\tabularnewline
\hline 
\end{tabular}\caption{Half width of the normalized profile in units of $z_{0}$ for different
values of $L_{q\bar{q}}$.\label{tab:Flux-profile}}
\end{table}
In Table(\ref{tab:Flux-profile}) we summarize the relationship between
the thickness and length of a classical flux-tube. The table clearly indicates that as the distance between the quark and the anti-quark, $L_{q\bar{q}}$, is increased a flux-tube is formed with a width which is asymptotically independent of $L_{q\bar{q}}$.

\subsection{Flux-tube with fluctuating thickness\label{subsec:Fluctuation-intrinsicTHickness}}

\begin{figure}[h]
\includegraphics[scale=0.4]{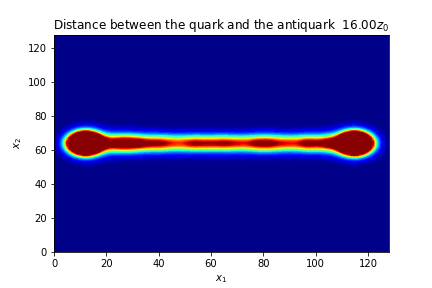}\includegraphics[scale=0.4]{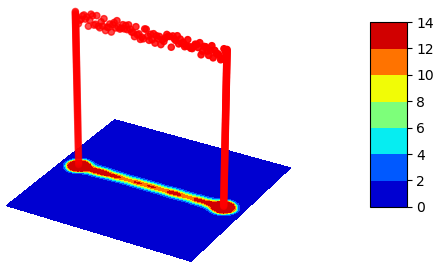}\caption{Static radially fluctuating open string and the corresponding flux-tube\label{fig:Fluctuation-of-intrinsic}}
\end{figure}

From the very inception of the gauge/gravity duality it was recognized
that the position of an object along the extra radial direction is
related to the size of it's holographic projection in the boundary
theory \cite{Susskind:1998dq}. For a confining theories this suggests
that the size of the flux-tube must be related to the position of
the corresponding open string in bulk. Further an open string with
a fluctuating radial coordinate along the extra dimension will correspond
to a flux-tube with varying thickness. We will verify that this is
indeed the case for $\Qcdthree$. We do so by considering an open
rectangular string, like the one we considered in section (\ref{subsec:Rectangular-Fluxtube}),
on which we superimpose random fluctuations along the $z$ direction
and then calculate the profile of the corresponding flux-tube. We
present the result in Fig.(\ref{fig:Fluctuation-of-intrinsic}). 

\section{Discussion\label{sec:Discussion}}

It is natural to ask, is there any thing that we can learn about real
QCD from the toy example considered in the present work. There are
few ways in which the phenomena that we have explored can get reflected
in QCD. Firstly QCD flux-tubes should have an intrinsic thickness
which can fluctuate. These radial fluctuations should lead to an attractive
Yukawa like potential \cite{Kinar:1999xu} and it might be interesting to incorporate these fluctuations in the effective string models of the fragmentation of hadrons \cite{BERENYI2015210}.

The shape of
the flux-tubes (see Figs.(\ref{fig:Emerg-FLux-Tube}, \ref{fig:Long-flux-tube-created})
and Table (\ref{tab:Flux-profile})) suggests that in an effective
string description of QCD flux-tube the string tension should depend
on the distance between the quark and the anti-quark and approaches
its constant value only as that distance tends to infinity \cite{Nielsen:1973cs,Vyas:2010uq}.
This would imply that when the distance between the quark and the
anti-quark is of the order of the confinement scale then there should
be deviations from the heavy quark potential calculated using an infinitely
long open string \cite{Aharony:2010db,Aharony:2010cx}. For in this
regime the open string in the gravitational description is not an
infinitely long open string placed at the confinement scale but has
to descent to the conformal boundary. In the boundary theory this
is reflected in the fact that when the quark anti-quark distance is
of the order of the confinement scale then the size of the blob of
action density surrounding the quark and the antiquark is comparable
to the distance between the quark and the anti-quark. A preliminary
estimate of this effect was made in \cite{Vyas:2012fk} but a more
precise calculation and its comparison with a high precision lattice
simulations like \cite{Brandt:2017aa} should be revealing. 

Another place where the intrinsic shape of a confining flux-tube can
have a physical consequences is in the possible presence of a long
range spin-spin interaction term in the heavy quark potential. In
\cite{Kogut:1981gm} it was argued, using the analogy with $U(1)$
gauge theories, that the quantum fluctuations of an electric flux-tube
will induce a magnetic field which will couple the spins of the quark
and the anti-quark leading to a spin-spin dependent potential term
which falls as the fifth power of the distance between the quark and
the anti-quark. In \cite{Vyas:2007bi} this was analysed using an effective
string description for the expectation value of a spin-half Wilson Loop. It was found that
for an effective string with Dirichlet boundary conditions there is no spin-spin dependent potential term, but when an allowance was made for the intrinsic thickness of the flux-tube near the end points, by calculating the spin-string interaction not at the end points but at a finite distance away from the end points, then a spin-spin interaction was indeed obtained. The shape of the confining flux-tube that we have obtained in the present investigation gives some credence to the analysis of \cite{Vyas:2007bi}, for the flux-tubes that are induced in the boundary theory do not terminate at the position of the quark or the anti-quark, rather they extend beyond it. We do not know how to incorporate this feature of the confining flux-tubes in an effective string description (the analysis of \cite{Vyas:2007bi} is physically motivated but ad-hoc), perhaps the analysis of boundary conditions in the context of holographic stringy description of hadrons \cite{Sonnenschein:2016aa, Sonnenschein:2017ylo}, and the analysis of the boundary conditions of an open string in the bulk under holographic renormalization group  \cite{Casalderrey-Solana:2019vnc} may provide us with some guidance.


\acknowledgments

I am extremely grateful to Ofer Aharony for his very useful comments
on a draft version of this paper. I would also like to thank Gunnar Bali for his comments on the present version of the paper. I have benefited from my discussions with Adi Armoni, Prem Kumar and Carlos Nunez while visiting Particle Physics and Cosmology Group at Swansea University and would like to thank them. During the course of this work I have enjoyed the hospitality
of the Department of Physics and Astronomy, Aarhus University, and
the Theoretical Physics Group at Raman Research Institute, Bangalore.
At Aarhus University I would like to thank Steen Hannestad and at
RRI Madhavan Varadarajan. The present work was made possible
by the continual support and encouragement of Janaki Abraham.

\appendix

\section{Shape of the open string sourcing the dilaton field\label{sec:Shape-of-the}}

The shape of the classical open string connecting the quark to the
anti-quark can be obtained from 
\begin{equation}
S_{NG}[X\left(\sigma\right)]=\frac{1}{2\pi\alpha'}\int d^{2}\sigma\sqrt{g_{E}\left(X\right)}.\label{eq:NGaction}
\end{equation}
We parameterize the world-sheet as 
\begin{equation}
X\left(t,x\right):=\left(t,x_{1}=x,x_{2},y,Z\left(x\right)\right).\label{eq:WorldSheetParameterization}
\end{equation}
Using the metric (\ref{eq:confingMetricQCD3}) we obtain the induced world-sheet
metric for an open string lying in the $x_{1}-z$ plane as 
\begin{equation}
g_{ab}=\text{diag}\left\{ \left(\frac{R}{Z}\right)^{2},\left(\frac{R}{Z}\right)^{2}\left(1+\frac{1}{f\left(z\right)}\left(\frac{dZ}{dx}\right)^{2}\right)\right\} ,\label{eq:InducedWorldSheetMetric}
\end{equation}
where 
\begin{equation}
f\left(z\right)=\left(1-\left(\frac{z}{z_{0}}\right)^{4}\right).\label{eq:Deff(z)}
\end{equation}
The parametrized Nambu-Goto action then takes the form 
\begin{equation}
S_{NG}\left[X\right]=\frac{R^{2}}{2\pi\alpha'}\int dtdx\left\{ \frac{1}{Z\left(x\right)^{2}}\left(1+\frac{1}{f\left(z\right)}\left(\frac{dZ}{dx}\right)^{2}\right)^{1/2}\right\} .\label{eq:NGactionInStaticGauge}
\end{equation}
Following \cite{Kinar:1998vq} we note that the action does not depend explicitly on $t$ therefore
for the time interval $\left(-T/2,T/2\right)$ it can be written as
\[
S_{NG}^{0}=\int_{-L/2}^{L/2}dx_{1}L\left(Z,Z'\right),
\]
where 
\begin{equation}
L\left(Z,Z'\right)=C_{0}\frac{1}{Z^{2}}\left(1+\frac{1}{f\left(z\right)}\left(\frac{dZ}{dx_{1}}\right)^{2}\right)^{1/2}\label{eq:L-staticOpenString-1}
\end{equation}
and we have defined the constant 
\begin{equation}
C_{0}=\frac{TR^{2}}{2\pi\alpha'}.\label{eq:defC0-1}
\end{equation}
$L$ describes a one-dimensional problem with $x_{1}$ playing the
role of time parameter, since $L$ does not depend explicitly on $x_{1}$
there is a conserved quantity 
\begin{equation}
H=P\frac{dZ}{dx_{1}}-L=PZ'-L,\label{eq:defHoneDimension-1}
\end{equation}
where the momentum conjugate to $Z$ is given by 
\begin{equation}
P=\frac{\partial L}{\partial\left(\frac{dZ}{dx_{1}}\right)}=\frac{\partial L}{\partial Z'}.\label{eq:defP-1}
\end{equation}
Using the above equations we obtain 
\begin{equation}
H=\frac{-C_{0}}{Z^{2}\left(1+\frac{1}{f\left(Z\right)}\left(\frac{dZ}{dx_{1}}\right)^{2}\right)^{1/2}},\label{eq:conservedH-1}
\end{equation}
which is conserved 
\begin{equation}
\frac{dH}{dx_{1}}=0.\label{eq:conservationOfH-1}
\end{equation}
By symmetry 
\begin{equation}
\left(\frac{dZ}{dx_{1}}\right)_{x_{1}=0}=0,\label{eq:Z'atx1=00003D00003D0-1}
\end{equation}
let 
\begin{equation}
Z\left(0\right)=z_{m}.\label{eq:defzm-1}
\end{equation}
Conservation of $H$ allows us to write 
\begin{equation}
\frac{1}{Z^{2}\left(1+\frac{1}{f\left(Z\right)}\left(\frac{dZ}{dx_{1}}\right)^{2}\right)^{1/2}}=\frac{1}{z_{m}^{2}},\label{eq:valueOfH-1}
\end{equation}
and we can obtain an equation for $Z\left(x_{1}\right)$ 
\begin{equation}
\frac{dZ}{dx_{1}}=\pm\sqrt{f\left(z\right)\left(\left(\frac{z_{m}}{Z}\right)^{4}-1\right)}\label{eq:shapeZ(x1)-1}
\end{equation}
the $\pm$ signs corresponding to the left and the right part of the
string that connects a quark at $-L/2$ to an anti-quark at $L/2$.
Using above result we can calculate 
\begin{equation}
\sqrt{g_{E}\left(Z_{c}\right)}=\frac{1}{Z\left(x\right)^{2}}\left(1+\frac{1}{f\left(z\right)}\left(\frac{dZ}{dx}\right)^{2}\right)^{1/2}=R^{2}\frac{z_{m}^{2}}{Z\left(x\right)^{4}}\label{eq:sqrt_gE}
\end{equation}

Next consider the positive branch of (\ref{eq:shapeZ(x1)-1}), and
noting that $Z\left(x_{1}\right)$ is an invertible function, we obtain
\begin{equation}
\frac{dX_{1}}{dz}=\left(f\left(z\right)\left(\left(\frac{z_{m}}{Z}\right)^{4}-1\right)\right)^{-1/2},\label{eq:X(z)forTheOpenString}
\end{equation}
which is more convenient for numerical calculations.

\section{Dilaton source term due to a rectangular string\label{subsec:Dilaton-Rect-Source}}

The dilaton string coupling for a static string configuration is given
by 
\begin{equation}
S_{ds}\left[X\left(\sigma\right)\right]=\frac{1}{4\pi l_{s}^{2}}\int dtd\sigma\Phi\left[X\left(\sigma\right)\right]\sqrt{g_{E}\left(X\left(\sigma\right)\right)}.\label{eq:stringDilatonAction}
\end{equation}
For a rectangular source of a dilaton field (\ref{fig:Open-string-Rectange})
we will parametrize the two sides of the rectangle by $\sigma=z,$and
obtain 
\begin{equation}
S_{\text{Rect. side}}=\frac{\lambda^{1/2}}{4\pi}\int dtdz\Phi\left[X\left(z\right)\right]\frac{1}{z^{2}\left(1-z^{4}\right)^{1/2}},\label{eq:recSource-Sides}
\end{equation}
with 
\begin{equation}
X\left(\sigma\right):=\left(t,x_{1q},0,0,z\right).\label{eq:sideStringProfile}
\end{equation}
For the top of the rectangle we use $\sigma=x_{1}$ as the parameter,
\begin{equation}
S_{\text{Rect. top}}=\frac{\lambda^{1/2}}{4\pi}\int dtdx_{1}\Phi\left[X\left(x_{1}\right)\right]\frac{1}{Z^{2}}\left(1+\frac{1}{\left(1-Z^{4}\right)}\left(\frac{dZ}{dx_{1}}\right)^{2}\right)^{1/2},\label{eq:srsRectTop}
\end{equation}
where the string profile is given by 
\begin{equation}
X\left(\sigma\right):=\left(t,x_{1},0,0,Z\left(x_{1}\right)\right).\label{eq:topStringProfile}
\end{equation}

\section{Numerical solution of the dilaton field equation\label{sec:Appendix:-Numerical-solution}}



The discrete version of the dilaton field equation (\ref{eq:discreteDilatonFieldEquation})
can be solved using relaxation algorithm \cite{koonin1998computational}

\begin{multline}
\Phi(m_{1},m_{2},m_{3})\rightarrow\left(1-\omega\right)\Phi\left(m_{1},m_{2},m_{3}\right)\\
+\frac{\omega}{C\left(m_{3}\right)}(A_{m_{3}+1/2}(\Phi\left(m_{1}-1,m_{2},m_{3}\right)+\Phi\left(m_{1}+1,m_{2},m_{3}\right)\\
+\Phi\left(m_{1},m_{2}-1,m_{3}\right)+\Phi\left(m_{1},m_{2}+1,m_{3}\right))\\
+\frac{\omega}{C\left(m_{3}\right)}\left(\sigma^{2}B_{m_{3}-1/2}\Phi\left(m_{2},m_{3}-1\right)+\sigma^{2}B_{m_{3}+1/2}^{2}\Phi\left(m_{2},m_{3}+1\right)\right)\\
-\frac{\omega}{C\left(m_{3}\right)}\frac{\sigma z_{m}^{2}\lambda^{1/2}}{8\pi}\left(\delta\left(0,m_{2}\right)\delta\left(Z_{c}\left(m_{1}\right),m_{3}\right)\frac{1}{Z_{c}^{4}\left(m_{1}\right)}\right),\label{eq:relaxationFor3DimCase}
\end{multline}
with $C\left(m_{3}\right)$ defined as

\[
C\left(m_{3}\right)=4A+\sigma^{2}B_{m_{3}-1/2}+\sigma^{2}B_{m_{3}+1/2},
\]
and $\omega$ is the relaxation parameter.


For the two dimensional case discussed in sec.(\ref{sec:Intrinsic-Thicknes-of})
the discrete version of the dilaton field equation (\ref{eq:twoDimDiscreteDilatonFieldEquation})
can be solved by the following relaxation procedure \cite{koonin1998computational},
setting $\sigma=1,$ 
\begin{multline}
\Phi(m_{2},m_{3})=\left(1-\omega\right)\Phi\left(m_{2},m_{3}\right)\\
+\frac{\omega}{C\left(m_{3}\right)}\left(A_{m_{3}+1/2}\left(\Phi\left(m_{2}-1,m_{3}\right)+\Phi\left(m_{2}+1,m_{3}\right)\right)\right)\\
+\frac{\omega}{C\left(m_{3}\right)}\left(B_{m_{3}-1/2}\Phi\left(m_{2},m_{3}-1\right)+B_{m_{3}+1/2}\Phi\left(m_{2},m_{3}+1\right)\right)\\
-\frac{\omega}{C\left(m_{3}\right)}\left(\frac{\lambda^{1/2}}{8\pi}\frac{\delta_{0,m_{2}}\delta_{N_{m},m_{3}}}{Z_{\text{lattice}}^{2}\left(0\right)}\right),\label{eq:relaxationFor2DimCase}
\end{multline}
where 
\[
C\left(m_{3}\right)=2A_{m_{3}+1/2}+B_{m_{3}-1/2}+B_{m_{3}+1/2}
\]
and $\omega$ is again the relaxation parameter.


In using relaxation algorithm one encounters the problem of critical
slowing down, namely that the long wavelength part of the solution
takes very large number of iterations to relax to its true value,
the number of iterations diverging as the size of the lattice increases.
For example if we consider a linear problem 
\begin{equation}
L\mathbf{u}=\mathbf{s},\label{eq:linear system}
\end{equation}
with a given source vector $\mathbf{s}$ and tries to solve for the
vector $\mathbf{u}$ using a relaxation algorithm then one finds that
the norm of the \emph{residue }vector 
\begin{equation}
\mathbf{r}=\mathbf{s}-L\mathbf{u},\label{eq:residue}
\end{equation}
initially decreases with the number of iterations but after a while
it stops reducing. To mitigate the problem of critical slowing down
we used multigrid algorithm \cite{trottenberg2001multigrid,stewart2017python}.

For the case of three dimensional anisotropic lattice we used the
following definition of the norm

\begin{equation}
\left\Vert \mathbf{r}\right\Vert =\sqrt{\left(a*b^{2}\mathbf{r}\cdot\mathbf{r}\right)},\label{eq:normResidue-3dim}
\end{equation}
where $a$ is the lattice constant along the $z$ direction while
$b$ is the lattice constant along $x_{1}$ and $x_{2}$ direction.
For the two dimensional case with an isotropic lattice constant $a$
the norm was taken to be 
\begin{equation}
\left\Vert \mathbf{r}\right\Vert =\sqrt{\left(a^{2}\mathbf{r}\cdot\mathbf{r}\right)}\label{eq:normResidue-2dim}
\end{equation}


All the numerical calculations involved in this work were done using
python 3.0 JupyterLab notebooks which are included as supplemental
material. Easiest way to run these notebooks is to install free and
open-source python distribution like \href{https://www.anaconda.com/distribution/}{Anaconda}.
The four notebooks included as ancillary files in the ArXiv submission are: 
\begin{enumerate}
\item \emph{ShapeOfFluxTube-1.ipynb} which was used for the calculations
reported in section (\ref{subsec:Emergence-of-fluxtube}). This notebook
depends on two python files 
\begin{enumerate}
\item \emph{utility\_3DimMG.py} 
\item \emph{AiStringSource.py} 
\end{enumerate}
\item \emph{RectangularString.ipynb} which was used for calculations reported
in section (\ref{subsec:Rectangular-Fluxtube}). This notebook depends
on the python file 
\begin{enumerate}
\item \emph{utility\_3DimMG.py} 
\end{enumerate}
\item \emph{ShapeOfFluxTube-3.ipynb }which was used for calculations reported
in section (\ref{subsec:Numerical-Infinite-String}). This notebook
depends on the python file 
\begin{enumerate}
\item \emph{utility\_2DimMG.py} 
\end{enumerate}
\item \emph{flucRectString.ipynb }which was used for the calculations reported
in section (\ref{subsec:Fluctuation-intrinsicTHickness}).This notebook
depends on the python file 
\begin{enumerate}
\item \emph{utility\_2DimMG.py} 
\end{enumerate}
\end{enumerate}

\end{document}